\documentclass[10pt,letterpaper,twocolumn]{article} %% two column, final layout

\usepackage{ol2}
\usepackage[draft]{hyperref}
\usepackage{amsmath}

\begin{document}

\twocolumn[ %% activate for two-column option

\title{Nonreciprocity in giant Goos-H\"anchen shift due to symmetry breaking}

%%%%%%%% For REVTeX it is possible to automate superscript and e-mail callouts with the superscriptaddress option; see REVTeX4 documentation.

\author{Madhuri Kumari and S Dutta Gupta$^*$}

\address{School of Physics, University of Hyderabad, Hyderabad 500046, India\\
$^*$Corresponding author: sdghyderabad@gmail.com
}

\begin{abstract}
We study giant Goos-H\"anchen (GH) shift in reflection from a near-symmetric coupled waveguide structure. We show that broken spatial symmetry can lead to GH shift with different signs for illumination from the opposite ends, a  direct consequence of the nonreciprocity relations considered earlier (Opt. Lett. 27,1205 (2002)). Symmetry breaking by adding a few nm thick bio-layer to one of the guides is enough to observe the nonreciprocity. This may have far reaching implications for efficient biosensing.
\end{abstract}


 ] %% activate for two-column option

\noindent Reciprocity relations and their violation have been one of the central themes of physics irrespective of the specifics of any given area \cite{nonreciprocity}. In general they are closely linked to the time reversal and space inversion symmetries. In the context of a lossless stratified medium with broken spatial symmetry, intensity reflection is known to be reciprocal \cite{sdgnr}. However, the phases of the reflection coefficient under illumination from opposite ends, can exhibit altogether different behavior leading to subluminal and superluminal reflected pulses at the opposite ends \cite{sdgnr1}. In the framework of a stationary phase approximation, the delay or advancement of the pulse can be estimated by the Wigner phase time $\tau$, which is equal to the frequency derivative of the phase of the reflection coefficient $r$ ($\tau=\frac{\partial \phi_r}{\partial \omega}$)\cite{wigner}. From a different angle, under the same stationary phase approximation the derivative of the phase of the reflection coefficient with respect to the surface component of the wave vector yields the Artmann formula for Goos-H\"anchen shift ($D=-\frac{\partial \phi_r}{\partial k_{||}}$) \cite{goos, Artmann}. It is thus difficult to miss out the analogy between the two different phenomena, namely the Wigner delay and the GH shift. Both are due to finite extent of the relevant processes, the former in time domain while the latter in the space domain. It is also expected that the nonreciprocity in the phase of the reflection coefficient for a system with broken spatial symmetry will translate into the nonreciprocal GH shifts in such structures. The literature on GH shift is now truly vast marking the tremendous progress of research in this area. Complex set up for the measurement of the GH shift for a single act of total internal reflection \cite{fabien} is now simplified considerably \cite{cylprism}. A broader perspective on GH shift  now encompasses structures with multiple interfaces, thus allowing for Fabry-Perot, surface and guided modes\cite{opposite1,oppositegh,composite1,kivshar,opposite2,opposite3,spbad}. It also addresses materials ranging from dielectrics and metals to composites and metamaterials \cite{composite1,kivshar}. Some very recent work focuses on the quantum GH shift in graphene \cite{graphene}. Multilayered media supporting various resonances can lead to giant GH shifts with magnitude, comparable to the beam waist.  Both positive and negative GH shift for TE and TM polarization have been reported by many \cite{opposite1,opposite2,opposite3}. GH shift with both the signs for the same polarization has also been proposed \cite{oppositegh}. The only (to the best of our knowledge) report on nonreciprocity in GH shift is for reflection off antiferromagnets in presence of an external magnetic field \cite{lima}. In this letter we demonstrate the close link between nonreciprocity in GH shift with broken symmetry in the context of a coupled waveguide system, loaded on both sides by identical high index prisms. We show that symmetry breaking can lead to giant GH shifts with different signs for illumination from the opposite ends. Such a structure with minor modifications can be an ideal candidate for accurate displacement sensing and bio-sensing. Note that  GH shift has been used as an effective tool for high resolution sensing in recent years\cite{sensor1,sensor2,sensor3}.
%%%%%%%%%%%%%
\begin{figure}[htb]
\centerline{\includegraphics[width=6.0cm]{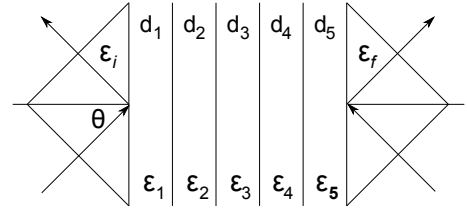}}
\caption{\label{fig1}Schematic view of the layered medium.}
\end{figure}
%%%%%%%%%%%%%
%%%%%%%%%%%%%%%%%%%%%%%%%%%%%%%%%%%%%%%
\begin{figure*}[t]
\centerline{\includegraphics[width=12cm]{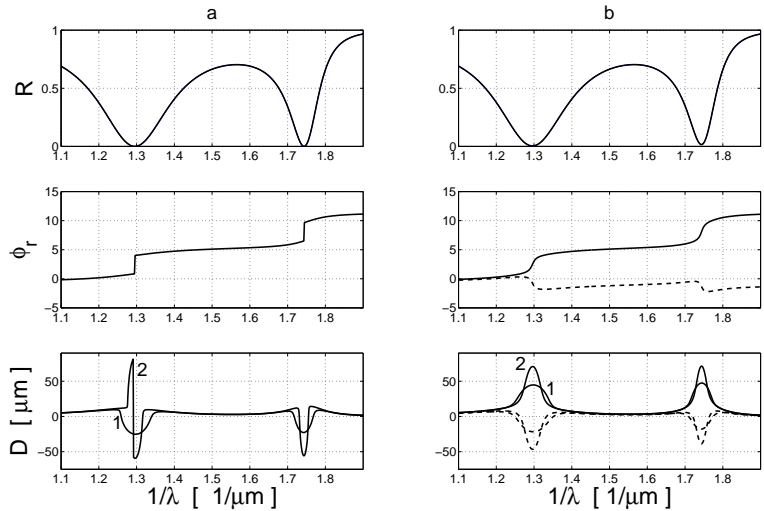}}
\caption{\label{fig2}(a) Intensity reflection coefficient $R=|r|^2$, phase $\phi_{R}$ of the reflection coefficient and the GH shift $D$ for a symmetric coupled waveguide structure as functions of inverse wavelength $1/\lambda$ for $d_{1}=d_{5}=d=0.1~\mu m,~d_{2}=d_{4}=0.3201~\mu m,~d_{3}=0.01~\mu m$. (b) Same as Fig. \ref{fig2}(a) except for broken symmetry with $d_{1}=d-0.005~\mu m,~d_{5}=d+0.005~\mu m$. The solid (dashed) line is for forward (backward) illumination. Curve marked by 1 (2) in the bottom row is for $w_{0}=50~\mu m$ ($w_{0}=100~\mu m$ ). Other parameters are as in the text.} 
\end{figure*}
%%%%%%%%%%%%%%%%%%%%%%%%%%%%%%%%%%%%%%
\par
Consider the resonant tunneling (RT) structure \cite{dheeraj} with two coupled waveguides as shown in Fig. \ref{fig1}. The motivation for choosing a coupled guided structure is to have extra sharp resonances and the associated enhancement in the GH shift. The choice of the RT configuration is dictated by two factors. It enables us to start with a symmetric structure with $d_{1}=d_{5}=d$, $d_{2}=d_{4}$, $\epsilon_{1}=\epsilon_{5}$ and $\epsilon_{2}=\epsilon_{4}$. Symmetry breaking can be achieved by shifting the coupled guides to, say, left by a small distance $s$ making $d_{1}=d-s$ and $d_{5}=d+s$,  leaving the total optical path intact. The second motivation is the ability of the structure to exhibit the signature of the coupled mode resonances in both reflection and transmission. Here we monitor only the attenuated reflection for $s$-polarized light. Analogous studies can easily be carried out for $p$-polarized light. Since we shall be comparing the two cases of illumination from left and from right, we label the former (latter) as forward (backward) denoted by subscript $f$ ($b$). 
\par
In order to calculate the GH shift we use the space frequency decomposition (SFD) method for a fundamental $s$-polarized Gaussian beam incident  at an angle $\theta$ \cite{fabien,carniglia}. Since many of the papers use the stationary phase (SP) method even for resonant structures, we compare the two at the end and bring out the limitations of the SP method. The reflection coefficient for the layered medium was calculated using the characteristics matrix method \cite{born}. Since the mathematical tools in both the above cases are well known, we do not repeat them here. The intensity reflection coefficient $R=|r|^{2}$, phase $\phi_{r}$ of the reflection coefficient $r$ and the GH shift $D$ for a lossless symmetric structure are shown in Fig. \ref{fig2}(a) as functions of inverse wavelength $1/\lambda$. The GH shift is plotted for two different beam waists, namely, $w_0=50 ~\mu m$ and $100 ~\mu m$. The parameters for computations were chosen as follows, $\epsilon_{i}=\epsilon_{f}=6.145$,  $\epsilon_{1}=\epsilon_{3}=\epsilon_{5}=2.2883$, $\epsilon_{2}=\epsilon_{4}=3.9085$, $d_{1}=d_{5}=d=0.1~\mu m$, $d_{2}=d_{4}=0.3201~\mu m$, $d_{3}=0.1~\mu m$, $\theta=48.76^0$. The width $d_{3}$ of the coupling layer controls the separation between the split modes shown in the top panel of Fig. \ref{fig2}(a). At each split resonance, as expected, the phase undergoes a sharp jump (middle pane) leading  to the enhanced GH shift (bottom panel). Needless to mention that because of the inherent symmetry one has the same response irrespective of the direction of illumination.  Note also that because of the singularity in the phase, the GH shift does not follow the Artmann formula and it can be partly positive or negative depending on the waist. The scenario changes drastically when the symmetry is disturbed by a tiny (5 nm) displacement of the coupled guides, say, to the left ($d_{1}=d-0.005~ \mu m ,~ d_{5}=d+0.005~ \mu m$). Due to the lossless character of the structure, time reversal symmetry is intact and one has the same response for $R$ for forward as well as backward illumination. As stated earlier, broken symmetry shows up in the nonreciprocity in phases (middle pane of Fig. \ref{fig2}(b)) with continuous derivative. The positive and the negative smooth steps in the phase result in GH shift with different signs for the two opposite directions (see bottom panel of Fig. \ref{fig2}(b)) yielding qualitatively analogous results as predicted by the Artmann formula. 
\par 
%%%%%%%%%%%%%%%%%%%%%%%%%%%%%%%%%%%%%%%%%%%%%%%%

%%%%%%%%%%%%%%%%
\begin{figure}[htb]
\centerline{\includegraphics[width=8.0cm]{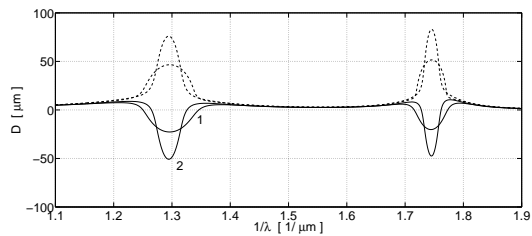}}
\caption{\label{fig3} GH shift with an additional 4 nm layer of Haemoglobin in between layer 1 and the left guide for forward (solid lines) and backward (dashed lines) illumination. Curve marked by 1 (2) is for $w_0=50~\mu m$ ($w_0=100~\mu m$). Other parameters are as in Fig. \ref{fig2}(a).}
\end{figure}
%%%%%%%%%%%%%%%%
\par
A close inspection of the bottom panel of Fig.2b reveals the remarkable possibility of such near- or pseudo- symmetric structures for high resolution sensing applications. The mere fact that such structures can exhibit opposite signs of shift for illumination from opposite sides immediately leads to nearly two-fold enhancement of the sensitivity. Referring to Fig. \ref{fig2}(b) a tiny displacement of 5 nm can cause a difference shifts ($|D_f-D_b|$) of about 110 $\mu m$ for a beam with 100 $\mu m$ waist. 
\par
We now demonstrate that the symmetry breaking by other means can also lead to similar effects. We insert a thin layer ($d_{12}=4~nm$) of Haemoglobin in between layer 1 and the left guide (see Fig. \ref{fig1}). The dielectric properties of the Haemoglobin is modelled by a resonant response, though our resonances are away from the Haemoglobin resonances in order to avoid losses. The results for forward and backward scattering for $w_{0}=50 ~\mu m$ and 100 $\mu m$ are shown in Fig. \ref{fig3}. It is clear from Fig. \ref{fig3} that such a nano-layer can easily be sensed by the huge difference shift of about 120 $\mu m$ for a 100 $\mu m$ beam. We also studied the effect of losses on the GH shift (not shown) by introducing a finite imaginary part to the dielectric response of the guides. Our studies reveal that the shift is extremely robust against losses  and the opposite signs of the GH shift (for the studied parameter ranges) persist. The main effect of the losses is a slight broadening, displacement and reduction of magnitude in the GH shift. 
\par
\begin{figure}[htb]
\centerline{\includegraphics[width=8.0cm]{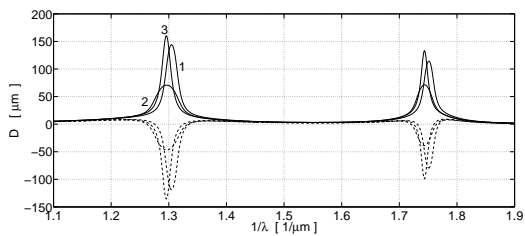}}
\caption{\label{fig4}Comparison of the GH shift calculated by SP and SFD methods. Solid(dashed) lines are for forward (backward) illumination. Curves labelled by 1,2 and 3 are for SP, SFD with $w_{0}=100~\mu m$ and $w_{0}=500~\mu m$, respectively. Other parameters are as in Fig. \ref{fig2}(b)}
\end{figure}

 
 It should be noted that the SP approximation can be a poor approximation near resonances, though many of the authors use it to analyse resonant structures \cite{sensor1,sensor2}. This is mainly because of the  drastic change in the phase of the reflection coefficient at the resonance. The limitations of the SP method has been nicely noted by  Broe and Keller \cite{oppositegh}, who showed that the major contribution to the GH shift comes from the poles of the reflection coefficient. For the standard GH shift for total internal reflection, an increasing $w_{0}$ leads one closer to the plane wave limit, asymptotically approaching the Artmann formula, though with decreasing relative shift $D/w_0$. We now show that the same is not true for resonant structures as in our case. In our case  a broad beam can lead to shifts larger than predicted by the SP method. This is shown in Fig. \ref{fig4}, where we compared the SP with SFD results with two different beam waists. Note also that the relative shift $D/w_{0}$ decreases with broader beams. In fact in the plane wave limit the relative shift approaches zero.
%%%%%%%%%%%%%%%%%%%%%%%%%%%%%%%%%%%%%%%
%%%%%%%%%%%%%%%%%%%%%%%%%%%%%%%%%%%%%%%%%%%%%%%%

\par
In conclusion we have studied the effects of symmetry breaking on the Goos-H\"anchen shift from a coupled waveguide system under illumination from opposite ends. We have demonstrated a nonreciprocity resulting in opposite signs of GH shift. The giant opposite GH shift is shown to have tremendous potentials for detecting displacements of a few nanometers and in general for bio-sensing.
\par
One of the authors is grateful to Girish S Agarwal for suggestions related to the organization of the manuscript. The authors are thankful to Shourya for providing us with the Haemoglobin dielectric response. Financial help from the Department of Science and Technology, Government of India is thankfully acknowledged.
%%%%%%%%%%%%%%%%%%%%%%%%

\pagebreak

\end{document}